\documentclass[preprint,aip]{revtex4-1}
\usepackage{amsmath}
\usepackage{amssymb}
\usepackage{graphicx,color}

\begin{document}
%%%%%%%%%%%%%%%%%%%%%%%%%%%%%%%%%%%%%%%%%%%%%%%%%%%%%%%%%%%%%%%%%%%%%
\title{Influences of quantum mechanically mixed electronic and vibrational pigment states in 2D electronic spectra of photosynthetic systems: Strong electronic coupling cases}

%\author{Yuta Fujihashi}
%\affiliation{Institute for Molecular Science, National Institutes of Natural Sciences, Okazaki 444-8585, Japan}

%\author{Graham R. Fleming}
%\affiliation{Department of Chemistry, University of California, Berkeley and Physical Biosciences Division, Lawrence Berkeley National Laboratory, Berkeley, CA 94720, USA}

%\author{Akihito Ishizaki}
%\email{ishizaki@ims.ac.jp}
%\affiliation{Institute for Molecular Science, National Institutes of Natural Sciences, Okazaki 444-8585, Japan}

\author{Yuta Fujihashi}
\affiliation{Institute for Molecular Science, National Institutes of Natural Sciences, Okazaki 444-8585, Japan}

\author{Graham R. Fleming}
\affiliation{Department of Chemistry, University of California, Berkeley and Physical Biosciences Division, Lawrence Berkeley National Laboratory, Berkeley, CA 94720, USA}

\author{Akihito Ishizaki}
\email{ishizaki@ims.ac.jp}
\affiliation{Institute for Molecular Science, National Institutes of Natural Sciences, Okazaki 444-8585, Japan}

%\phone{+81 (0)564 55 7310}
%  \fax{+81 (0)564 53 4660}

%\abbreviations{IR,NMR,UV}
%\keywords{Photosynthetic energy transfer, dissipative quantum dynamics, 2D electronic spectra}

%%%%%%%%%%%%%%%%%%%%%%%%%%%%%%%%%%%%%%%%%%%%%%%%%%%%%%%%%%%%%%%%%%%%%
\begin{abstract}
In 2D electronic spectroscopy studies, long-lived quantum beats have recently been observed in photosynthetic systems, and several theoretical studies have suggested that the beats are produced by quantum mechanically mixed electronic and vibrational states. Concerning the electronic-vibrational quantum mixtures, the impact of protein-induced fluctuations was examined by calculating the 2D electronic spectra of a weakly coupled dimer with the Franck-Condon active vibrational modes in the resonant condition [Fujihashi {et al.}, {\it J. Chem. Phys.} {\bf 142}, 212403 (2015), arXiv:1505.05281.]. 
This analysis demonstrated that quantum mixtures of the vibronic resonance are rather robust under the influence of the fluctuations at cryogenic temperatures, whereas the mixtures are eradicated by the fluctuations at physiological temperatures. 
However, this conclusion cannot be generalized because the magnitude of the coupling inducing the quantum mixtures is proportional to the inter-pigment electronic coupling. 
In this study, we explore the impact of the fluctuations on electronic-vibrational quantum mixtures in a strongly coupled dimer with an off-resonant vibrational mode. Toward this end, we calculate energy transfer dynamics and 2D electronic spectra of a model dimer that corresponds to the most strongly coupled bacteriochlorophyll molecules in the Fenna-Matthews-Olson complex in a numerically accurate manner. The quantum mixtures are found to be robust under the exposure of protein-induced fluctuations at cryogenic temperatures, irrespective of the resonance. At 300\,K, however, the quantum mixing is disturbed more strongly by the fluctuations, and therefore, the beats in the 2D spectra become obscure even in a strongly coupled dimer with a resonant vibrational mode.
Further, the overall behaviors of the energy transfer dynamics are demonstrated to be dominated by the environment and coupling between the $0-0$ vibronic transitions as long as the Huang-Rhys factor of the vibrational mode is small. The electronic-vibrational quantum mixtures do not necessarily play a significant role in electronic energy transfer dynamics despite contributing to the enhancement of long-lived quantum beating in the 2D spectra.
\end{abstract}
\maketitle

%%%%%%%%%%%%%%%%%%%%%%%%%%%%%%%%%%%%%%%%%%%%%%%%%%%%%%%%%%%%%%%%%%%%%
\section{Introduction}

The development of ultrafast nonlinear spectroscopic techniques has provided an incisive tool to probe the quantum dynamics of condensed phase systems.
\cite{Mukamel:1995us}
Investigating photosynthetic pigment-protein complexes using two-dimensional (2D) electronic spectroscopy revealed long-lived beating phenomena in the 2D spectra \cite{Engel:2007hb, Calhoun:2009bn, Collini:2010fy, Panitchayangkoon:2010fw,Panitchayangkoon:2011cs, SchlauCohen:2012dn, Westenhoff:2012fi,Dawlaty:2012fs,Fuller:2014iz,Romero:2014jm}. Engel et al. \cite{Engel:2007hb} revealed the presence of quantum beats persisting for at least 660\,fs in the 2D electronic spectra of the Fenna-Matthews-Olson (FMO) pigment-protein complex from green sulfur bacteria at cryogenic temperatures. Panitchayangkoon et al. \cite{Panitchayangkoon:2010fw} subsequently demonstrated that quantum beats in the FMO complex persist for at least 1.5\,ps and 300\,fs at cryogenic and physiological temperatures, respectively. Recently, nuclear vibrational effects have been theorized to explain the spectral beating behavior \cite{Turner:2011ef,Christensson:2011ht, Christensson:2012gp, Kolli:2012ip,YuenZhou:2012hu, CaycedoSoler:2012ib, Butkus:2012hn, Butkus:2013fy, Tiwari:2013dt, Kreisbeck:2013jva, Chin:2013ia, Tiwari:2013dt, Chenu:2013cf, Plenio:2013bg,Rivera:2013kb,Seibt:2013dp,Halpin:2014jd, Tempelaar:2014vu,OReilly:2014it,Perlik:2014bd}, which persist for significantly longer than the predicted electronic dephasing times.\cite{Ishizaki:2009jg,Ishizaki:2009ky}
Christensson et al. \cite{Christensson:2012gp} proposed that the resonance between electronic states and the Franck-Condon active vibrational states serves to create vibronic excitons, i.e., quantum mechanically mixed electronic and vibrational states.
Such states have vibrational characteristics and enhanced transition dipole moments owing to intensity borrowing from strong electronic transitions, despite the small Huang-Rhys factors of bacteriochlorophyll (BChl).\cite{Wendling:2000ha,Ratsep:2007fq,Ratsep:2011cq,Schulze:2014iv} Therefore, vibronic resonance was suggested to produce oscillations in the 2D signal that persist for several picoseconds. In this regard, Tiwari, Peters, and Jonas \cite{Tiwari:2013dt} showed that vibronic resonance also enhances the vibrational coherence in the electronic ground state and demonstrated that this effect could explain the long-lived oscillations in the 2D spectra.

Indeed, the concept of vibronic resonance plausibly explains the spectroscopically observed long-lived spectral beating. However, the dynamic interplay between the electronic-vibrational quantum mixture and electronic energy fluctuations induced by the environmental dynamics naturally raises a question. \cite{Ishizaki:2010ft,Ishizaki:2013jg}
To address this question, Fujihashi, Fleming, and Ishizaki \cite{Fujihashi:2015kz} considered a weakly coupled model dimer \cite{Chenu:2013cf} inspired by BChls 3 and 4 in the FMO complex of {\it Chlorobaculum tepidum} ({\it C. tepidum}) \cite{vanAmerongen:2000fy,Cho:2005bm,Adolphs:2006ey}, in which the gap between the two electronic energy eigenstates resonates with one of the strongest vibrational modes with frequency $180\,{\rm cm^{-1}}$ in the absence of the surrounding environment.
They numerically explored the impacts of fluctuations on the quantum mixture of the vibronic resonance by analyzing the beating amplitudes of the ground state bleaching contribution in calculated nonrephasing 2D spectra. They demonstrated that the vibronic resonance is rather robust under the influence of environmentally induced fluctuations at cryogenic temperatures, resulting in long-lasting beats of vibrational origin in the 2D spectra. However, larger-amplitude fluctuations disturbed the quantum mixture of the vibronic resonance more strongly at physiological temperatures, causing the beats in the 2D spectra to disappear. Further, they suggested that vibronic resonance does not play a significant role in the electronic energy transfer (EET) dynamics, at least in the weakly coupled dimer, despite contributing to the enhancement of the long-lived quantum beating in the 2D electronic spectra.

However, this conclusion cannot be generalized. Here, it should be noted that the magnitudes of the couplings inducing electronic-vibrational quantum mixtures are dependent on inter-pigment electronic couplings as well as the Huang-Rhys factors of the vibrational modes. Indeed, there exist several pairs of BChls constituting strongly coupled dimers in the FMO complex. Even if the gaps between the electronic energy eigenstates of such strongly coupled dimers do not resonate with relatively strong vibrational modes in BChl molecules, it may be possible to induce relevant electronic-vibrational quantum mixtures that affect the time evolution of the 2D spectra and/or corresponding EET dynamics. This study mainly aims to explore the influences of quantum mechanically mixed electronic and vibrational states in a strongly coupled model dimer under the exposure of protein-induced fluctuations. Toward this end, we study the 2D electronic spectra and EET dynamics in a model dimer mimicking BChls 1 and 2 of the FMO complex, which is the most strongly coupled pair.

%%%%%%%%%%%%%%%%%%%%%%%%%%%%%%%%%%%%%%%%
\section{Theory}
\setcounter{equation}{0}

We consider a coupled hetero-dimer, the simplest EET system. To describe the EET dynamics and optical responses, we restrict the electronic spectra of the $m$-th pigment in a pigment-protein complex (PPC) to the ground state, $\lvert \varphi_{mg} \rangle$, and the first excited state, $\lvert \varphi_{me} \rangle$, although higher excited states are occasionally of consequence in nonlinear spectroscopic signals.\cite{Brixner:2004ko,Bruggemann:2007gu} No experimental evidence has been reported for nonadiabatic transitions and radiative/nonradiative decays between $\lvert \varphi_{me} \rangle$ and $\lvert \varphi_{mg} \rangle$ in light-harvesting complexes on picosecond timescales. Thus, we organize the product states in the order of their elementary excitation numbers. The overall ground state with zero excitation reads $\lvert g \rangle = \lvert \varphi_{1g} \rangle \lvert \varphi_{2g} \rangle$. The presence of a single excitation at pigment 1 is expressed as $\lvert 1 \rangle = \lvert \varphi_{1e}\rangle\lvert \varphi_{2g}\rangle$ and the other, as $\lvert 2 \rangle = \lvert \varphi_{1g}\rangle\lvert \varphi_{2e}\rangle$. As the intensity of sunlight is weak, the single-excitation manifold is of primary importance under physiological conditions. However, nonlinear spectroscopic techniques such as 2D electronic spectroscopy can populate some higher excitation manifolds, e.g., the double-excitation manifold comprising $\lvert 12 \rangle = \lvert \varphi_{1e}\rangle\lvert \varphi_{2e}\rangle$. Therefore, the PPC Hamiltonian can be expressed as
\begin{align}
	\hat{H}_{\rm PPC}
	=
	\hat{H}_{\rm exc} + \hat{H}_{\rm env} + \hat{H}_{\rm exc-env},
	\label{decomposed-hamiltonian}
\end{align}
The first part $\hat{H}_{\rm exc}$ is the Hamiltonian describing the electronic transition expressed as
\begin{align}
	\hat{H}_{\rm exc}
	&=
	E_g \lvert g \rangle\langle g \rvert
	\notag\\
	&+
	\sum_{m=1}^2 (E_g + \hbar\Omega_m)
	\lvert m \rangle \langle m \rvert
	+
	\sum_{m=1}^2 \sum_{n \ne m}
	\hbar J_{mn}
	\lvert m \rangle\langle n \rvert
	\notag\\
	&+
	(E_g + \hbar\Omega_1 + \hbar\Omega_2)
	\lvert 12 \rangle \langle 12 \rvert,
\end{align}
where $\hbar\Omega_m$ is the Franck-Condon transition energy of the $m$th pigment, and $\hbar J_{mn}$ denotes the electronic coupling between the $m$th and the $n$th pigments. In the derivations above, the energy of the overall electronic ground state, $E_g$, has been introduced; however, we set $E_g = 0$ without loss of generality. The second term $\hat{H}_{\rm env}$ is the Hamiltonian of the environmental and vibrational DOFs. The last term $\hat{H}_{\rm exc-env}$ describes the coupling of the environmental and vibrational degrees of freedom (DOFs) to the electronic excitations as
\begin{align}
	\hat{H}_{\rm exc-env}
	&=
	\sum_{m=1}^2 \hat{u}_m \lvert m \rangle \langle m \rvert
	+
	(\hat{u}_1 + \hat{u}_2) \lvert 12 \rangle \langle 12 \rvert.
\end{align}
The operator $\hat{u}_m$ is termed the collective energy gap coordinate to describe the fluctuations in the electronic energy and the dissipation of the reorganization energy. In this study, we assumed that the fluctuations in the electronic energies of different pigments are not correlated.
It should be emphasized that the coordinate, $\hat{u}_m$, includes information associated with not only the electronic excited state but also the electronic ground state, as demonstrated in Ref.~\cite{Fujihashi:2015kz}.
The electronic coupling between the pigments may also depend on the environmental and nuclear vibrational DOFs. However, in this study, we assume that the nuclear dependence of $\hbar J_{mn}$ is vanishingly small, and we employ the Condon-like approximation as usual.

We assume that the environmentally induced fluctuations can be described as Gaussian processes and that the relevant nuclear dynamics are harmonic vibrations. Under this assumption, the dynamic properties of the environmental and intramolecular vibrational motions are characterized by the two-body correlation functions of $\hat{u}_m(t) = e^{i\hat{H}_{\rm env}t/\hbar} \hat{u}_m e^{-i\hat{H}_{\rm env}t/\hbar}$ as
\begin{align}
	\langle \hat{u}_m(t) \hat{u}_m(0) \rangle
	=
	D_m(t) - i\frac{\hbar}{2} \Phi_m(t),
	\label{2-body-correlation}
\end{align}
where the canonical average has been introduced, $\langle \dots \rangle = \mathrm{Tr}(\dots e^{-\beta\hat{H}_{\rm env}})/\mathrm{Tr}\,e^{-\beta\hat{H}_{\rm env}}$, with $\beta$ being the inverse temperature, $1/k_{\rm B}T$. In Eq.~\eqref{2-body-correlation}, $D_m(t)$ is the symmetrized correlation function of $\hat{u}_m(t)$, defined as $D_m(t) = (1/2) \left\langle \{ \hat{u}_m(t), \hat{u}_m(0) \}_+ \right\rangle$, and $\Phi_m(t)$ is the response function defined by $\Phi_m(t) = (i/\hbar)\left\langle [ \hat{u}_m(t), \hat{u}_m(0) ]_- \right\rangle$. The quantum fluctuation-dissipation relation \cite{Kubo:1985bs} allows one to express the symmetrized correlation and response functions as
\begin{gather}
	D_m(t) = \frac{\hbar}{\pi}\int^\infty_0 d\omega J_m(\omega) \coth\frac{\beta\hbar\omega}{2}\cos\omega t,
	\label{symmetrized-correlation}
	\\
	\Phi_m(t) = \frac{2}{\pi}\int^\infty_0 d\omega J_m(\omega) \sin\omega t,
\end{gather}
where $J_m(\omega)$ is termed the spectral density and is defined by the imaginary part of the Fourier-Laplace transform of the response function. The absolute magnitude of the spectral density is related to the total reorganization energy, $\hbar\lambda_{m,{\rm env}} + \hbar\lambda_{m,{\rm vib}} = \int^\infty_0 d\omega J_m(\omega)/(\pi\omega)$, where $\hbar\lambda_{m,{\rm env}}$ and $\hbar\lambda_{m,{\rm vib}}$ denote the environmental and vibrational reorganization energies of the $m$th pigment, respectively. The spectral density can be separated into two-components: a continuous part corresponding to the environmental contribution and discrete peaks corresponding to the vibrational modes, namely, \cite{Adolphs:2006ey, Womick:2011iw, Kolli:2012ip, Chin:2013ia, Fujihashi:2015kz}
\begin{align}
	J_m(\omega)
	=
	J_{m,{\rm env}}(\omega) + J_{m,{\rm vib}}(\omega).
\end{align}
To focus on the timescale of the environmental dynamics affecting the electronic transition energies, we model the environmental component as the Drude-Lorentz spectral density as
\begin{align}
	J_{m, {\rm env}}(\omega) = 2\hbar\lambda_{\rm env}\frac{\gamma_{\rm env}\omega}{\omega^2 + \gamma_{\rm env}^2},
	\label{env-spectral-density}
\end{align}
where $\gamma_{\rm env}^{-1}$ corresponds to the timescale of the environment-induced fluctuations.
For simplicity, we consider a single intramolecular vibration on each pigments with frequency $\omega_{\rm vib}$ and Huang-Rhys factor $S$. We describe the spectral density for the vibrational mode using the Brownian oscillator model with the vibrational relaxation rate $\gamma_{\rm vib}$, such that
\begin{align}
	J_{m,{\rm vib}}(\omega)
	=
	\hbar\lambda_{\rm vib}
	\frac{\omega_{\rm vib}^2}{\tilde\omega_{\rm vib}}
	\left[
		\frac{\gamma_{\rm vib}}{(\omega-\tilde\omega_{\rm vib})^2+\gamma_{\rm vib}^2}
		-
		\frac{\gamma_{\rm vib}}{(\omega+\tilde\omega_{\rm vib})^2+\gamma_{\rm vib}^2}
	\right],
	\label{vib-spectral-density}
\end{align}
where $\tilde\omega_{\rm vib}=(\omega_{\rm vib}^2-\gamma_{\rm vib}^2)^{1/2}$ and $\hbar\lambda_{\rm vib}=\hbar\omega_{\rm vib}S$.
In the slow vibrational relaxation limit ($\gamma_{\rm vib}\to 0$), Eq.~\eqref{vib-spectral-density} is reduced to $J_{m,{\rm vib}}(\omega) = \pi \hbar S \omega^2 [\delta(\omega-\omega_{\rm vib})-\delta(\omega+\omega_{\rm vib})]$, which has been employed in the literature.\cite{Adolphs:2006ey, Womick:2011iw}

An adequate description of the EET dynamics is given by the reduced density operator, $\hat\rho_{\rm exc}(t)$, that is, the partial trace of the total PPC density operator, $\hat\rho_{\rm PPC}(t)$, over the environmental and nuclear DOFs: $\hat\rho_{\rm exc}(t) = \mathrm{Tr}_{\rm env}\hat\rho_{\rm PPC}(t)$.
For this reduction, we assume that the total PPC system at the initial time ($t=0$) is in the factorized product state of the form $\hat\rho_{\rm PPC}^{\rm eq}=\hat\rho_{\rm exc} \otimes e^{-\beta\hat{H}_{\rm env}}/\mathrm{Tr}\,e^{-\beta\hat{H}_{\rm env}}$.
In other words, $\hat\rho_{\rm exc}(t) = \mathrm{Tr}_{\rm env}[\hat{G}(t) \hat\rho_{\rm PPC}^{\rm eq} ]$, where $\hat{G}(t)$ is the retarded propagator of the total PPC system in the Liouville space.
This factorized initial condition is generally considered to be unphysical in the literature on open quantum systems as it neglects the inherent correlation between a system of interest and its environment. \cite{Grabert:1988ha,Tanimura:2014er} However, in electronic excitation processes, this initial condition is inconsequential because it corresponds to the electronic ground state or electronic excited state generated in accordance with the vertical Franck-Condon transition.\cite{Ishizaki:2009jg,Ishizaki:2010fx}
The Gaussian property of $\hat{u}_m$ enables one to derive a formally exact equation of motion that can describe the EET dynamics under the influence of the environmentally induced fluctuation and dissipation. Therefore, the EET dynamics under the influence of the environmental/nuclear dynamics and corresponding 2D electronic spectra can be described in a numerically accurate manner. The technical details of the computations are given in Ref. \cite{Fujihashi:2015kz}.

%%%%%%%%%%%%%%%%%%%%%%%%%%%%%%%%%%%%%%%%
\section{Results and discussion}

In this section, we present and discuss the numerical results to explore the influences of quantum mechanically mixed electronic and vibrational states in a strongly coupled dimer under the exposure of protein-induced fluctuations. Toward this end, we address a model dimer mimicking BChls 1 and 2 in the FMO complex of {\it C. tepidum} \cite{vanAmerongen:2000fy,Cho:2005bm,Adolphs:2006ey,Chenu:2013cf}; this is the most strongly coupled pair and exhibits wave-like EET persisting for several hundred femtoseconds even in the absence of vibronic contributions.\cite{Ishizaki:2009ky} The Franck-Condon transition energies of BChls 1 and 2 are set to $\Omega_1=12410\,{\rm cm^{-1}}$ and $ \Omega_2 = \Omega_1 + 120\,{\rm cm^{-1}}$, and their electronic coupling is set to $J_{12} = -87.7\,{\rm cm^{-1}}$ according to Ref.~\cite{Adolphs:2006ey}.
The transition dipole moment directions are set to $\theta_{12} = 66.6^\circ$ based on the PDB file 3ENI.\cite{Tronrud:2009ch}

In the descriptions below, $\lvert \chi_{av}^m \rangle$ denotes the $v$-th vibrational level of the $a$-th electronic state in the $m$-th pigment. For simplicity, we consider a single vibrational mode with frequency $\omega_{\rm vib}$ of BChl 1 only, and we focus on the quantum mixture between the vibronic transition, $\lvert\varphi_{1g}\rangle \lvert \chi_{g0}^1\rangle \leftrightarrow \lvert\varphi_{1e}\rangle \lvert \chi_{e1}^{1}\rangle$, of BChl 1 and the electronic transition, $\lvert \varphi_{2g}\rangle \leftrightarrow \lvert \varphi_{2e}\rangle$, of BChl 2.
The mixture is induced by the coupling $J_{12} \langle \chi_{e1}^1 \vert \chi_{g0}^1 \rangle = -J_{12}\exp(-S/2)\sqrt{S}$, where $S$ denotes the Huang-Rhys factor of the mode. The vibrational modes of BChl in solutions or protein environments have been investigated,\cite{Wendling:2000ha,Ratsep:2007fq,Ratsep:2011cq,Schulze:2014iv} and the strongest vibrational mode has been found at approximately $180\,{\rm cm^{-1}}$. However, this vibrational mode does not resonate with the gap between the electronic energy eigenstates in the model dimer under investigation, $[{(\Omega_2 -\Omega_1)^2 +4J_{12}^2}]^{1/2} = 213\,{\rm cm^{-1}}$. Therefore, we investigate two cases concerning the vibrational frequency: the off-resonance case, $\omega_{\rm vib}=180\,{\rm cm^{-1}}$, and the resonance case, $\omega_{\rm vib}=213\,{\rm cm^{-1}}$.

For the numerical calculations of the 2D spectra and EET dynamics, the environmental reorganization energy and reorganization time constant are set to $\lambda_{\rm env}=35{\rm \,cm^{-1}}$ and $\gamma_{\rm env}^{-1}=100\,{\rm fs}$, and the vibrational relaxation rate is set to $\gamma_{\rm vib}^{-1} = 2\,{\rm ps}$. The Huang-Rhys factors of BChl are typically small, \cite{Fransted:2012fs} and the factor associated with the vibrational mode in this work is set to $S=0.025$,\cite{Tiwari:2013dt} which is within the range of the experimentally measured values.\cite{Ratsep:2007fq,Ratsep:2011cq}

%%%%%%%%%%%
\subsection{Beating in 2D electronic spectra}

\begin{figure}%%%%%%%%%%
         \includegraphics{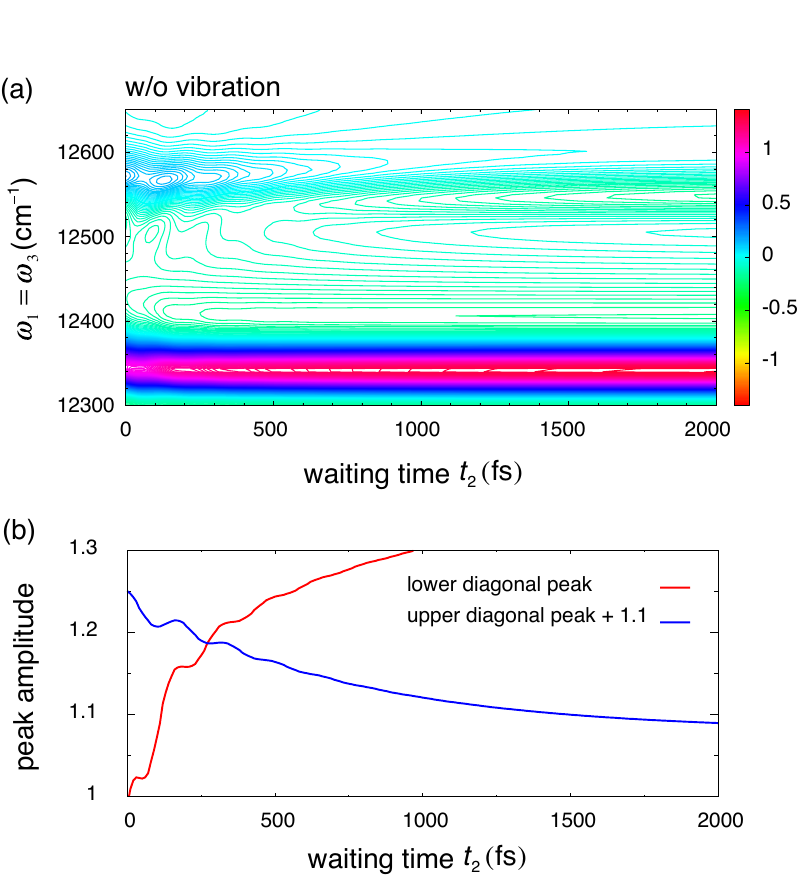}
	\caption{(a) Time evolution of the diagonal cut of the nonrephasing 2D spectra at 77 K of a coupled-dimer without intramolecular vibration. (b) Time evolutions of the amplitudes of the lower diagonal peak, $(\omega_1=\omega_3=12,343\,{\rm cm^{-1}})$, and the upper diagonal peak, $(\omega_1=\omega_3=12,582\,{\rm cm^{-1}})$, in the spectra of Fig.~\ref{fig:full-diagonal-J90-77K}a. The environmental reorganization energy and the reorganization time constant are set to $\lambda_{\rm env}=35\,{\rm cm^{-1}}$ and $\gamma_{\rm env}^{-1}=100\,{\rm fs}$, and the Huang-Rhys factor and the vibrational relaxation time constant are $S=0.025$ and $\gamma_{\rm vib}^{-1} = 2\,{\rm ps}$. The normalization of the plots is such that the maximum value of the nonrephasing 2D spectra of Fig.~\ref{fig:full-diagonal-J90-77K}a at $t_2=0\,{\rm fs}$ is unity, and equally spaced contour levels ($0, \pm 0.01, \pm 0.02, \dots, $) are drawn. To clarify the relationship between the beating amplitude magnitudes in Fig.~\ref{fig:full-diagonal-J90-77K}b, the peak amplitudes are shifted.}
	\label{fig:full-diagonal-J90-77K}
\end{figure}

\begin{figure}%%%%%%%%%%
         \includegraphics{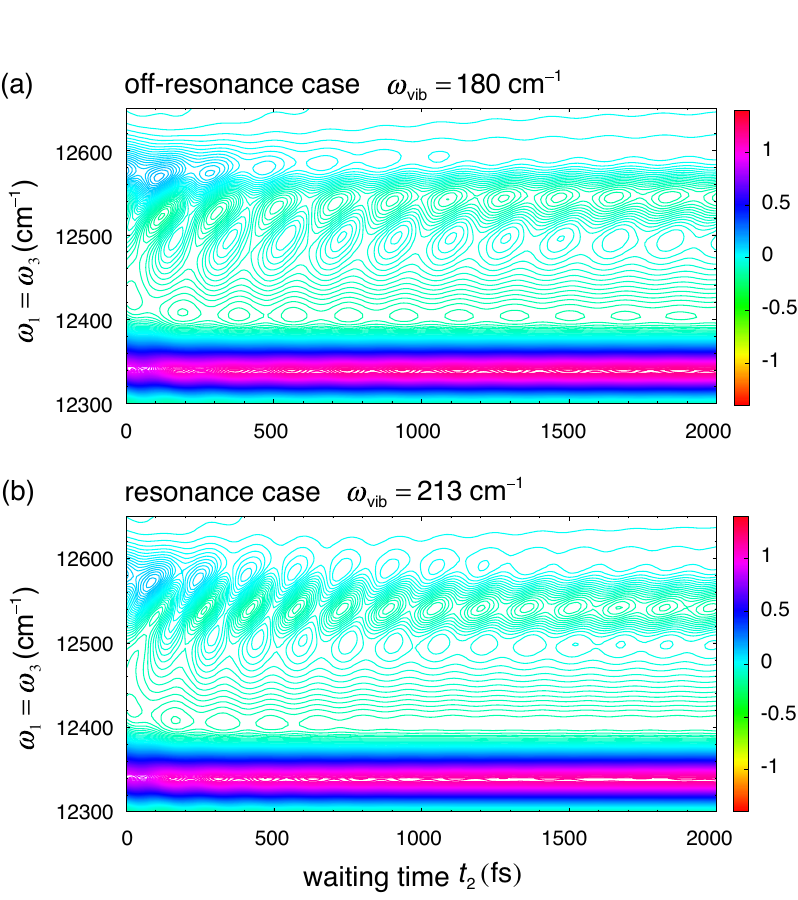}
	\caption{Time evolution of the diagonal cut of the nonrephasing 2D spectra at 77 K of a coupled-dimer with intramolecular vibration of frequency (a) $\omega_{\rm vib}=180\,{\rm cm^{-1}}$ and (b) $\omega_{\rm vib}=213\,{\rm cm^{-1}}$. The normalization of the plots is such that the maximum value of the nonrephasing 2D spectra in Fig.~\ref{fig:full-diagonal-J90-77K}a at $t_2=0\,{\rm fs}$ is unity, and equally spaced contour levels ($0, \pm 0.01, \pm 0.02, \dots, $) are drawn.}
	\label{fig:full-diagonal-J90-77K-2}
\end{figure}

In this subsection, we explore the contributions of the vibrational modes of the strongly coupled dimer in 2D electronic spectra. 
In Fig.~\ref{fig:full-diagonal-J90-77K}, we consider 2D spectra in the absence of any vibrational contributions as a reference. Figure~\ref{fig:full-diagonal-J90-77K}a shows the diagonal cut of the calculated nonrephasing 2D spectra as a function of the waiting time at 77 K, and Fig.~\ref{fig:full-diagonal-J90-77K}b shows a plot of the time evolution of the amplitudes of the diagonal peaks $(\omega_1=\omega_3=12343\,{\rm cm^{-1}}, 12582\,{\rm cm^{-1}})$.
The contour plots in Fig.~\ref{fig:full-diagonal-J90-77K}a apparently indicate the monotonous decay of the upper diagonal peak induced by the exciton relaxation as well as the monotonic increase of the lower diagonal peak, which is the inherent nature of nonrephasing signals irrespective of the EET dynamics. Nevertheless, Fig.~\ref{fig:full-diagonal-J90-77K}b reveals the presence of spectral beating persisting for 600 fs. This quantum beating originates from the strong electronic coupling between BChls 1 and 2 and is thus of electronic origin. Indeed, this electronic coherence and its lifetime are consistent with the previous theoretical analysis of EET dynamics in the FMO complex.\cite{Ishizaki:2009ky} Figure~\ref{fig:full-diagonal-J90-77K-2} shows the time evolutions of the calculated nonrephasing 2D spectra at 77 K with the vibrational mode of BChl 1 for (a) the off-resonance case, $\omega_{\rm vib}=180\,{\rm cm^{-1}}$, and (b) the resonance case, $\omega_{\rm vib}=213\,{\rm cm^{-1}}$. The spectral beats of the 2D spectra are enhanced by the vibrational contributions and persist for at least 2 ps. This observation demonstrates that even in the off-resonance case, the electronic-vibrational quantum mixture causes distinct beats in the 2D spectra of strongly coupled dimers at cryogenic temperatures.

\begin{figure}%%%%%%%%%%
         \includegraphics{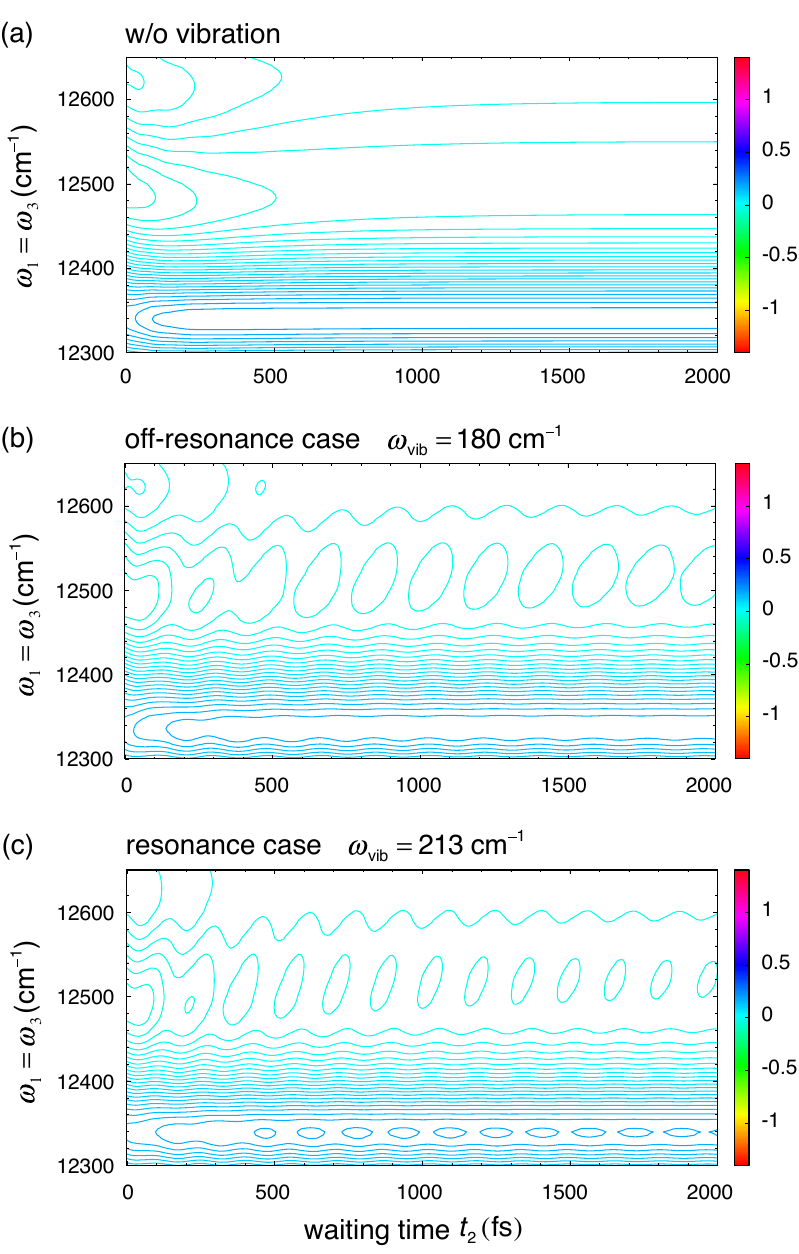}
	\caption{Time evolution of the diagonal cut of the nonrephasing 2D spectra of (a) a coupled-dimer without intramolecular vibration, (b) a coupled-dimer with intramolecular vibration of frequency $\omega_{\rm vib}=180\,{\rm cm^{-1}}$, and (c) a coupled-dimer with vibration of frequency $\omega_{\rm vib}=213\,{\rm cm^{-1}}$ at 300 K. The normalization of the plots is such that the maximum value of the nonrephasing 2D spectra in Fig.~\ref{fig:full-diagonal-J90-77K}a at $t_2=0\,{\rm fs}$ is unity, and equally spaced contour levels ($0, \pm 0.01, \pm 0.02, \dots, $) are drawn.}
         \label{fig:full-diagonal-J90-300K}	
\end{figure}

However, the variance of the amplitude of the environmentally induced fluctuations in the electronic energy is estimated via Eq.~\eqref{symmetrized-correlation} as $D_{m,{\rm env}}(0) \simeq 2k_{\rm B}T\int^\infty_0 d\omega {J_{m,{\rm env}}(\omega)}/{\pi\omega}$, where the high-temperature approximation, $\coth(\beta\hbar\omega/2) \simeq 2k_{\rm B}T/\hbar\omega$, has been employed.
The fluctuations at higher temperatures are larger in magnitude, and therefore, the electronic-vibrational quantum mixture is supposedly fragile at physiological temperatures.
Figure~\ref{fig:full-diagonal-J90-300K}a shows the time evolution of the diagonal cut of the calculated nonrephasing 2D spectra at 300 K in the absence of the vibrational contribution, and Figs.~\ref{fig:full-diagonal-J90-300K}b and \ref{fig:full-diagonal-J90-300K}c show the 2D spectra with the vibrational mode in the off-resonance case, $\omega_{\rm vib}=180\,{\rm cm^{-1}}$, and the resonance case, $\omega_{\rm vib}=213\,{\rm cm^{-1}}$, respectively.
In both the off-resonance and resonance cases, the amplitudes of the beating decrease significantly, although Figs.~\ref{fig:full-diagonal-J90-300K}b and \ref{fig:full-diagonal-J90-300K}c show spectral beating arising from the electronic-vibrational mixture, albeit only slightly. This indicates that the environmentally induced fluctuations at physiological temperatures almost destroy the quantum mixing even in a strongly coupled dimer.
In this study, inhomogeneous broadening in the 2D spectra that is caused by the static disorder of the electronic transition energies has not been considered. Under the influence of inhomogeneous broadening, as seen in the experimental data, the small amplitudes of the beats observed in Fig.~\ref{fig:full-diagonal-J90-300K} are presumably disturbed and are difficult to observe.

%%%%%%%%%%%
\subsection{Population dynamics}

\begin{figure}%%%%%%%%%%
	\includegraphics{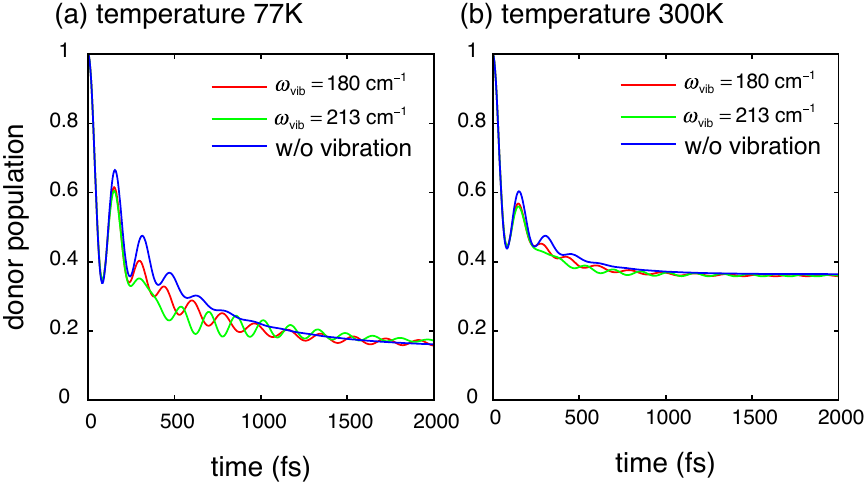}
	\caption{Time evolution of the energy donor population affected by intramolecular vibration of BChl 1 (acceptor) at temperatures of (a) 77 K and (b) 300 K. In each panel, the red and green lines indicate the dynamics in the presence of vibrations of frequency $\omega_{\rm vib}=180\,{\rm cm^{-1}}$ and $\omega_{\rm vib}=213\,{\rm cm^{-1}}$, respectively. The blue lines indicate the dynamics without the vibrations. The calculations were performed with the parameters employed in Fig.~\ref{fig:full-diagonal-J90-77K}.}
	\label{fig:pop-dynamics-env-strong}
\end{figure}

We explore the influence of intramolecular vibration on the spatial dynamics of electronic excitation. Figure~\ref{fig:pop-dynamics-env-strong} shows the time evolution of the energy donor (BChl 2) population affected by the intramolecular vibration at temperatures of (a) 77 K and (b) 300 K. The red and green lines indicate the dynamics in the presence of vibrations with frequency $\omega_{\rm vib}=180\,{\rm cm^{-1}}$ and $\omega_{\rm vib}=213\,{\rm cm^{-1}}$, respectively. The blue lines indicate the dynamics without the vibrations as a reference.
The EET dynamics in Fig.~\ref{fig:pop-dynamics-env-strong} show wave-like motion induced by the quantum superposition between the $0-0$ vibronic transition of BChl 1 and the electronic transition of BChl 2 because of the strong electronic coupling.
In the case without a vibrational mode, the electronic coherence persists for 700 fs and 300 fs at 77 K and 300 K, respectively.
The coupling constant inducing the electronic-vibrational quantum mixture $J_{12} \langle \chi_{e1}^1 \vert \chi_{g0}^1 \rangle $ is proportional to the electronic coupling.
Consequently, the EET dynamics in the case of strong electronic coupling shows long-lasting (up to at least 2 ps) wave-like motion induced by the quantum mechanically mixed electronic and vibrational states in contrast to that in the case of the small electronic coupling investigated in Ref.~\cite{Fujihashi:2015kz}.
The vibrational mode slightly contributes to the acceleration of the EET because the electronic-vibrational quantum mixture adds a new EET channel originating from the $0-1$ vibronic transition of BChl 1 and the electronic transition of BChl 2.
As the vibrational frequency approaches the resonance condition, the contribution of the vibrational mode is increased, as shown in Fig.~\ref{fig:pop-dynamics-env-strong}.
However, the overall behaviors of EET dynamics are dominated by the environment and the coupling between the $0-0$ vibronic transition of BChl 1 and the electronic transition of BChl 2. The inter-site coupling constant inducing the the quantum mixture, $J_{12} \langle \chi_{e1}^1 \vert \chi_{g0}^1 \rangle = -J_{12}\exp(-S/2)\sqrt{S} = 13.7\,{\rm cm^{-1}}$, is much smaller than the coupling constant between the $0-0$ vibronic transition of BChl 1 and the electronic transition of BChl 2, $J_{12} \langle \chi_{e0}^1 \vert \chi_{g0}^1 \rangle \simeq J_{12}\exp(-S/2) = - 86.6\,{\rm cm^{-1}}$, which gives rise to the fast channel, as long as the Huang-Rhys factor is small.
Consequently, the electronic-vibrational quantum mixture does not play a significant role in enhancing the EET dynamics under the influence of the environmentally induced fluctuations even in the most strongly coupled BChls in the FMO complex, despite contributing to the enhancement of the long-lived beating in the 2D electronic spectra.

%%%%%%%%%%%%%%%%%%%%%%%%%%%%%%%%%%%%%%%%
\section{Concluding Remarks}

In this study, we have explored the impact of environmentally induced fluctuations on the quantum mechanically mixed electronic and vibrational states of pigments in a strongly coupled dimer.
Toward this end, we investigated BChls 1 and 2, the most strongly coupled dimer in the FMO complex, and performed accurate numerical calculations of the EET dynamics and corresponding 2D electronic spectra.

In the strongly coupled dimer at cryogenic temperatures, the electronic-vibrational quantum mixing is rather robust against the environmentally induced fluctuations irrespective of the resonance, resulting in long-lasting spectral beating of vibrational origin in the 2D electronic spectra. However, the amplitudes of the environmentally induced fluctuations in the electronic excitation energy increase with the temperature.
At physiological temperatures, the electronic-vibrational quantum mixture is disturbed more strongly, and therefore, the beating in 2D spectra becomes obscure despite the strong electronic coupling.

Furthermore, we have investigated the effect of the underdamped intramolecular vibrational motion on the EET dynamics using the same equation of motion and parameters as in the calculations of the 2D electronic spectra.
The electronic-vibrational quantum mixture leads to an oscillation lasting up to several picoseconds at cryogenic and physiological temperatures.
However, the overall behavior of the EET dynamics is dominated by the environment and the coupling between the $0-0$ vibronic transition of BChl 1 and the electronic transition of BChl 2 as long as the Huang-Rhys factor is small. The electronic-vibrational quantum mixture contributes only slightly to the acceleration of the electronic energy transfer, although the 2D electronic spectra clearly exhibit long-lived beating enhanced by the quantum mixture.
This suggests that information about the oscillatory behavior extracted from the 2D spectra should be interpreted carefully.

%%%%%%%%%%%%%%%%%%%%%%%%%%%%%%%%%%%%%%%%%%%%%%%%%%%%%%%%%%%%%%%%%%%%%
\begin{acknowledgements}
This study was supported by a Grant-in-Aid for Scientific Research (No. 25708003) from the Japan Society for the Promotion of Science and the U.S. Department of Energy, Office of Science, Office of Basic Energy Sciences, Chemical Sciences, Geosciences, and Biosciences Division.
\end{acknowledgements}

\bibliography{./aigrp_bib}

%Merlin.mbs v4.21 2009-07-09.
\begin{thebibliography}{10}%
\makeatletter
\providecommand \@ifxundefined [1]{%
 \ifx #1\undefined \expandafter \@firstoftwo
 \else \expandafter \@secondoftwo
\fi
}%
\providecommand \@ifnum [1]{%
 \ifnum #1\expandafter \@firstoftwo
 \else \expandafter \@secondoftwo
\fi
}%
\providecommand \enquote [1]{``#1''}%
\providecommand \bibnamefont  [1]{#1}%
\providecommand \bibfnamefont [1]{#1}%
\providecommand \citenamefont [1]{#1}%
\providecommand\href[0]{\@sanitize\@href}%
\providecommand\@href[1]{\endgroup\@@startlink{#1}\endgroup\@@href}%
\providecommand\@@href[1]{#1\@@endlink}%
\providecommand \@sanitize [0]{\begingroup\catcode`\&12\catcode`\#12\relax}%
\@ifxundefined \pdfoutput {\@firstoftwo}{%
 \@ifnum{\z@=\pdfoutput}{\@firstoftwo}{\@secondoftwo}%
}{%
 \providecommand\@@startlink[1]{\leavevmode\special{html:<a href="#1">}}%
 \providecommand\@@endlink[0]{\special{html:</a>}}%
}{%
 \providecommand\@@startlink[1]{%
  \leavevmode
  \pdfstartlink
   attr{/Border[0 0 1 ]/H/I/C[0 1 1]}%
   user{/Subtype/Link/A<</Type/Action/S/URI/URI(#1)>>}%
  \relax
 }%
 \providecommand\@@endlink[0]{\pdfendlink}%
}%
\providecommand \url  [0]{\begingroup\@sanitize \@url }%
\providecommand \@url [1]{\endgroup\@href {#1}{\urlprefix}}%
\providecommand \urlprefix [0]{URL }%
\providecommand \Eprint[0]{\href }%
\@ifxundefined \urlstyle {%
  \providecommand \doi [1]{doi:\discretionary{}{}{}#1}%
}{%
  \providecommand \doi [0]{doi:\discretionary{}{}{}\begingroup
  \urlstyle{rm}\Url }%
}%
\providecommand \doibase [0]{http://dx.doi.org/}%
\providecommand \Doi[1]{\href{\doibase#1}}%
\providecommand \selectlanguage [0]{\@gobble}%
\providecommand \bibinfo [0]{\@secondoftwo}%
\providecommand \bibfield [0]{\@secondoftwo}%
\providecommand \translation [1]{[#1]}%
\providecommand \BibitemOpen[0]{}%
\providecommand \bibitemStop [0]{}%
\providecommand \bibitemNoStop [0]{.\EOS\space}%
\providecommand \EOS [0]{\spacefactor3000\relax}%
\providecommand \BibitemShut [1]{\csname bibitem#1\endcsname}%
%</preamble>
\bibitem{Mukamel:1995us}%
  \BibitemOpen
  \bibfield{author}{%
  \bibinfo {author} {\bibfnamefont{Shaul}\ \bibnamefont{Mukamel}},\ }%
  \emph{\bibinfo {title} {{Principles of Nonlinear Optical Spectroscopy}}}\
  (\bibinfo {publisher} {Oxford University Press},\ \bibinfo {address} {New
  York},\ \bibinfo {year} {1995})\BibitemShut{NoStop}%
\bibitem{Engel:2007hb}%
  \BibitemOpen
  \bibfield{author}{%
  \bibinfo {author} {\bibfnamefont{Gregory~S}\ \bibnamefont{Engel}}, \bibinfo
  {author} {\bibfnamefont{Tessa~R}\ \bibnamefont{Calhoun}}, \bibinfo {author}
  {\bibfnamefont{Elizabeth~L}\ \bibnamefont{Read}}, \bibinfo {author}
  {\bibfnamefont{Tae~Kyu}\ \bibnamefont{Ahn}}, \bibinfo {author}
  {\bibfnamefont{Tom{\'a}{\v s}}\ \bibnamefont{Man{\v c}al}}, \bibinfo {author}
  {\bibfnamefont{Yuan-Chung}\ \bibnamefont{Cheng}}, \bibinfo {author}
  {\bibfnamefont{Robert~E}\ \bibnamefont{Blankenship}},\ and\ \bibinfo {author}
  {\bibfnamefont{Graham~R}\ \bibnamefont{Fleming}},\ }%
  \bibfield{title}{%
  \enquote{\bibinfo {title} {{Evidence for wavelike energy transfer through
  quantum coherence in photosynthetic systems}},}\ }%
  \bibfield{journal}{%
  \bibinfo {journal} {Nature}\ }%
  \textbf{\bibinfo {volume} {446}},\ \bibinfo {pages} {782--786} (\bibinfo
  {year} {2007})\BibitemShut{NoStop}%
\bibitem{Calhoun:2009bn}%
  \BibitemOpen
  \bibfield{author}{%
  \bibinfo {author} {\bibfnamefont{Tessa~R}\ \bibnamefont{Calhoun}}, \bibinfo
  {author} {\bibfnamefont{Naomi~S}\ \bibnamefont{Ginsberg}}, \bibinfo {author}
  {\bibfnamefont{Gabriela~S}\ \bibnamefont{Schlau-Cohen}}, \bibinfo {author}
  {\bibfnamefont{Yuan-Chung}\ \bibnamefont{Cheng}}, \bibinfo {author}
  {\bibfnamefont{Matteo}\ \bibnamefont{Ballottari}}, \bibinfo {author}
  {\bibfnamefont{Roberto}\ \bibnamefont{Bassi}},\ and\ \bibinfo {author}
  {\bibfnamefont{Graham~R}\ \bibnamefont{Fleming}},\ }%
  \bibfield{title}{%
  \enquote{\bibinfo {title} {{Quantum Coherence Enabled Determination of the
  Energy Landscape in Light-Harvesting Complex II}},}\ }%
  \bibfield{journal}{%
  \bibinfo {journal} {J. Phys. Chem. B}\ }%
  \textbf{\bibinfo {volume} {113}},\ \bibinfo {pages} {16291--16295} (\bibinfo
  {year} {2009})\BibitemShut{NoStop}%
\bibitem{Collini:2010fy}%
  \BibitemOpen
  \bibfield{author}{%
  \bibinfo {author} {\bibfnamefont{Elisabetta}\ \bibnamefont{Collini}},
  \bibinfo {author} {\bibfnamefont{Cathy~Y}\ \bibnamefont{Wong}}, \bibinfo
  {author} {\bibfnamefont{Krystyna~E}\ \bibnamefont{Wilk}}, \bibinfo {author}
  {\bibfnamefont{Paul M~G}\ \bibnamefont{Curmi}}, \bibinfo {author}
  {\bibfnamefont{Paul}\ \bibnamefont{Brumer}},\ and\ \bibinfo {author}
  {\bibfnamefont{Gregory~D}\ \bibnamefont{Scholes}},\ }%
  \bibfield{title}{%
  \enquote{\bibinfo {title} {{Coherently wired light-harvesting in
  photosynthetic marine algae at ambient temperature}},}\ }%
  \bibfield{journal}{%
  \bibinfo {journal} {Nature}\ }%
  \textbf{\bibinfo {volume} {463}},\ \bibinfo {pages} {644--647} (\bibinfo
  {year} {2010})\BibitemShut{NoStop}%
\bibitem{Panitchayangkoon:2010fw}%
  \BibitemOpen
  \bibfield{author}{%
  \bibinfo {author} {\bibfnamefont{G}~\bibnamefont{Panitchayangkoon}}, \bibinfo
  {author} {\bibfnamefont{D}~\bibnamefont{Hayes}}, \bibinfo {author}
  {\bibfnamefont{K~A}\ \bibnamefont{Fransted}}, \bibinfo {author}
  {\bibfnamefont{J~R}\ \bibnamefont{Caram}}, \bibinfo {author}
  {\bibfnamefont{E}~\bibnamefont{Harel}}, \bibinfo {author}
  {\bibfnamefont{J}~\bibnamefont{Wen}}, \bibinfo {author} {\bibfnamefont{R~E}\
  \bibnamefont{Blankenship}},\ and\ \bibinfo {author}
  {\bibfnamefont{Gregory~S}\ \bibnamefont{Engel}},\ }%
  \bibfield{title}{%
  \enquote{\bibinfo {title} {{Long-lived quantum coherence in photosynthetic
  complexes at physiological temperature}},}\ }%
  \bibfield{journal}{%
  \bibinfo {journal} {Proc. Natl. Acad. Sci. USA}\ }%
  \textbf{\bibinfo {volume} {107}},\ \bibinfo {pages} {12766--12770} (\bibinfo
  {year} {2010})\BibitemShut{NoStop}%
\bibitem{Panitchayangkoon:2011cs}%
  \BibitemOpen
  \bibfield{author}{%
  \bibinfo {author} {\bibfnamefont{G}~\bibnamefont{Panitchayangkoon}}, \bibinfo
  {author} {\bibfnamefont{D~V}\ \bibnamefont{Voronine}}, \bibinfo {author}
  {\bibfnamefont{D}~\bibnamefont{Abramavicius}}, \bibinfo {author}
  {\bibfnamefont{J~R}\ \bibnamefont{Caram}}, \bibinfo {author}
  {\bibfnamefont{N~H~C}\ \bibnamefont{Lewis}}, \bibinfo {author}
  {\bibfnamefont{S}~\bibnamefont{Mukamel}},\ and\ \bibinfo {author}
  {\bibfnamefont{Gregory~S}\ \bibnamefont{Engel}},\ }%
  \bibfield{title}{%
  \enquote{\bibinfo {title} {{Direct evidence of quantum transport in
  photosynthetic light-harvesting complexes}},}\ }%
  \bibfield{journal}{%
  \bibinfo {journal} {Proc. Natl. Acad. Sci. USA}\ }%
  \textbf{\bibinfo {volume} {108}},\ \bibinfo {pages} {20908--20912} (\bibinfo
  {year} {2011})\BibitemShut{NoStop}%
\bibitem{SchlauCohen:2012dn}%
  \BibitemOpen
  \bibfield{author}{%
  \bibinfo {author} {\bibfnamefont{Gabriela~S}\ \bibnamefont{Schlau-Cohen}},
  \bibinfo {author} {\bibfnamefont{Akihito}\ \bibnamefont{Ishizaki}}, \bibinfo
  {author} {\bibfnamefont{Tessa~R}\ \bibnamefont{Calhoun}}, \bibinfo {author}
  {\bibfnamefont{Naomi~S}\ \bibnamefont{Ginsberg}}, \bibinfo {author}
  {\bibfnamefont{Matteo}\ \bibnamefont{Ballottari}}, \bibinfo {author}
  {\bibfnamefont{Roberto}\ \bibnamefont{Bassi}},\ and\ \bibinfo {author}
  {\bibfnamefont{Graham~R}\ \bibnamefont{Fleming}},\ }%
  \bibfield{title}{%
  \enquote{\bibinfo {title} {{Elucidation of the timescales and origins of
  quantum electronic coherence in LHCII}},}\ }%
  \bibfield{journal}{%
  \bibinfo {journal} {Nat. Chem.}\ }%
  \textbf{\bibinfo {volume} {4}},\ \bibinfo {pages} {389--395} (\bibinfo {year}
  {2012})\BibitemShut{NoStop}%
\bibitem{Westenhoff:2012fi}%
  \BibitemOpen
  \bibfield{author}{%
  \bibinfo {author} {\bibfnamefont{Sebastian}\ \bibnamefont{Westenhoff}},
  \bibinfo {author} {\bibfnamefont{David}\ \bibnamefont{Pale\u{c}ek}}, \bibinfo
  {author} {\bibfnamefont{Petra}\ \bibnamefont{Edlund}}, \bibinfo {author}
  {\bibfnamefont{Philip}\ \bibnamefont{Smith}},\ and\ \bibinfo {author}
  {\bibfnamefont{Donatas}\ \bibnamefont{Zigmantas}},\ }%
  \bibfield{title}{%
  \enquote{\bibinfo {title} {{Coherent Picosecond Exciton Dynamics in a
  Photosynthetic Reaction Center}},}\ }%
  \bibfield{journal}{%
  \bibinfo {journal} {J. Am. Chem. Soc.}\ }%
  \textbf{\bibinfo {volume} {134}},\ \bibinfo {pages} {16484--16487} (\bibinfo
  {year} {2012})\BibitemShut{NoStop}%
\bibitem{Dawlaty:2012fs}%
  \BibitemOpen
  \bibfield{author}{%
  \bibinfo {author} {\bibfnamefont{Jahan~M}\ \bibnamefont{Dawlaty}}, \bibinfo
  {author} {\bibfnamefont{Akihito}\ \bibnamefont{Ishizaki}}, \bibinfo {author}
  {\bibfnamefont{Arijit~K}\ \bibnamefont{De}},\ and\ \bibinfo {author}
  {\bibfnamefont{Graham~R}\ \bibnamefont{Fleming}},\ }%
  \bibfield{title}{%
  \enquote{\bibinfo {title} {{Microscopic quantum coherence in a
  photosynthetic-light-harvesting antenna}},}\ }%
  \bibfield{journal}{%
  \bibinfo {journal} {Phil. Trans. R. Soc. A}\ }%
  \textbf{\bibinfo {volume} {370}},\ \bibinfo {pages} {3672--3691} (\bibinfo
  {year} {2012})\BibitemShut{NoStop}%
\bibitem{Fuller:2014iz}%
  \BibitemOpen
  \bibfield{author}{%
  \bibinfo {author} {\bibfnamefont{Franklin~D}\ \bibnamefont{Fuller}}, \bibinfo
  {author} {\bibfnamefont{Jie}\ \bibnamefont{Pan}}, \bibinfo {author}
  {\bibfnamefont{Andrius}\ \bibnamefont{Gelzinis}}, \bibinfo {author}
  {\bibfnamefont{Vytautas}\ \bibnamefont{Butkus}}, \bibinfo {author}
  {\bibfnamefont{S~Seckin}\ \bibnamefont{Senlik}}, \bibinfo {author}
  {\bibfnamefont{Daniel~E}\ \bibnamefont{Wilcox}}, \bibinfo {author}
  {\bibfnamefont{Charles~F}\ \bibnamefont{Yocum}}, \bibinfo {author}
  {\bibfnamefont{Leonas}\ \bibnamefont{Valkunas}}, \bibinfo {author}
  {\bibfnamefont{Darius}\ \bibnamefont{Abramavicius}},\ and\ \bibinfo {author}
  {\bibfnamefont{Jennifer~P}\ \bibnamefont{Ogilvie}},\ }%
  \bibfield{title}{%
  \enquote{\bibinfo {title} {{Vibronic coherence in oxygenic
  photosynthesis}},}\ }%
  \bibfield{journal}{%
  \bibinfo {journal} {Nat. Chem.}\ }%
  \textbf{\bibinfo {volume} {6}},\ \bibinfo {pages} {706--711} (\bibinfo {year}
  {2014})\BibitemShut{NoStop}%
\bibitem{Romero:2014jm}%
  \BibitemOpen
  \bibfield{author}{%
  \bibinfo {author} {\bibfnamefont{Elisabet}\ \bibnamefont{Romero}}, \bibinfo
  {author} {\bibfnamefont{Ramunas}\ \bibnamefont{Augulis}}, \bibinfo {author}
  {\bibfnamefont{Vladimir~I}\ \bibnamefont{Novoderezhkin}}, \bibinfo {author}
  {\bibfnamefont{Marco}\ \bibnamefont{Ferretti}}, \bibinfo {author}
  {\bibfnamefont{Jos}\ \bibnamefont{Thieme}}, \bibinfo {author}
  {\bibfnamefont{Donatas}\ \bibnamefont{Zigmantas}},\ and\ \bibinfo {author}
  {\bibfnamefont{Rienk}\ \bibnamefont{van Grondelle}},\ }%
  \bibfield{title}{%
  \enquote{\bibinfo {title} {{Quantum coherence in photosynthesis for efficient
  solar-energy conversion}},}\ }%
  \bibfield{journal}{%
  \bibinfo {journal} {Nat. Phys.}\ }%
  \textbf{\bibinfo {volume} {10}},\ \bibinfo {pages} {677--683} (\bibinfo
  {year} {2014})\BibitemShut{NoStop}%
\bibitem{Turner:2011ef}%
  \BibitemOpen
  \bibfield{author}{%
  \bibinfo {author} {\bibfnamefont{Daniel~B}\ \bibnamefont{Turner}}, \bibinfo
  {author} {\bibfnamefont{Krystyna~E}\ \bibnamefont{Wilk}}, \bibinfo {author}
  {\bibfnamefont{Paul M~G}\ \bibnamefont{Curmi}},\ and\ \bibinfo {author}
  {\bibfnamefont{Gregory~D}\ \bibnamefont{Scholes}},\ }%
  \bibfield{title}{%
  \enquote{\bibinfo {title} {{Comparison of Electronic and Vibrational
  Coherence Measured by Two-Dimensional Electronic Spectroscopy}},}\ }%
  \bibfield{journal}{%
  \bibinfo {journal} {J. Phys. Chem. Lett.}\ }%
  \textbf{\bibinfo {volume} {2}},\ \bibinfo {pages} {1904--1911} (\bibinfo
  {year} {2011})\BibitemShut{NoStop}%
\bibitem{Christensson:2011ht}%
  \BibitemOpen
  \bibfield{author}{%
  \bibinfo {author} {\bibfnamefont{N}~\bibnamefont{Christensson}}, \bibinfo
  {author} {\bibfnamefont{F}~\bibnamefont{Milota}}, \bibinfo {author}
  {\bibfnamefont{J}~\bibnamefont{Hauer}}, \bibinfo {author}
  {\bibfnamefont{J}~\bibnamefont{Sperling}}, \bibinfo {author}
  {\bibfnamefont{O}~\bibnamefont{Bixner}}, \bibinfo {author}
  {\bibfnamefont{A}~\bibnamefont{Nemeth}},\ and\ \bibinfo {author}
  {\bibfnamefont{H~F}\ \bibnamefont{Kauffmann}},\ }%
  \bibfield{title}{%
  \enquote{\bibinfo {title} {{High Frequency Vibrational Modulations in
  Two-Dimensional Electronic Spectra and Their Resemblance to Electronic
  Coherence Signatures}},}\ }%
  \bibfield{journal}{%
  \bibinfo {journal} {J. Phys. Chem. B}\ }%
  \textbf{\bibinfo {volume} {115}},\ \bibinfo {pages} {5383--5391} (\bibinfo
  {year} {2011})\BibitemShut{NoStop}%
\bibitem{Christensson:2012gp}%
  \BibitemOpen
  \bibfield{author}{%
  \bibinfo {author} {\bibfnamefont{Niklas}\ \bibnamefont{Christensson}},
  \bibinfo {author} {\bibfnamefont{Harald~F}\ \bibnamefont{Kauffmann}},
  \bibinfo {author} {\bibfnamefont{T{\~o}nu}\ \bibnamefont{Pullerits}},\ and\
  \bibinfo {author} {\bibfnamefont{Tom{\'a}{\v s}}\ \bibnamefont{Man{\v
  c}al}},\ }%
  \bibfield{title}{%
  \enquote{\bibinfo {title} {{Origin of Long-Lived Coherences in
  Light-Harvesting Complexes}},}\ }%
  \bibfield{journal}{%
  \bibinfo {journal} {J. Phys. Chem. B}\ }%
  \textbf{\bibinfo {volume} {116}},\ \bibinfo {pages} {7449--7454} (\bibinfo
  {year} {2012})\BibitemShut{NoStop}%
\bibitem{Kolli:2012ip}%
  \BibitemOpen
  \bibfield{author}{%
  \bibinfo {author} {\bibfnamefont{Avinash}\ \bibnamefont{Kolli}}, \bibinfo
  {author} {\bibfnamefont{Edward~J}\ \bibnamefont{O{\textquoteright}Reilly}},
  \bibinfo {author} {\bibfnamefont{Gregory~D}\ \bibnamefont{Scholes}},\ and\
  \bibinfo {author} {\bibfnamefont{Alexandra}\ \bibnamefont{Olaya-Castro}},\ }%
  \bibfield{title}{%
  \enquote{\bibinfo {title} {{The fundamental role of quantized vibrations in
  coherent light harvesting by cryptophyte algae}},}\ }%
  \bibfield{journal}{%
  \bibinfo {journal} {J. Chem. Phys.}\ }%
  \textbf{\bibinfo {volume} {137}},\ \bibinfo {pages} {174109} (\bibinfo {year}
  {2012})\BibitemShut{NoStop}%
\bibitem{YuenZhou:2012hu}%
  \BibitemOpen
  \bibfield{author}{%
  \bibinfo {author} {\bibfnamefont{Joel}\ \bibnamefont{Yuen-Zhou}}, \bibinfo
  {author} {\bibfnamefont{Jacob~J}\ \bibnamefont{Krich}},\ and\ \bibinfo
  {author} {\bibfnamefont{Al{\'a}n}\ \bibnamefont{Aspuru-Guzik}},\ }%
  \bibfield{title}{%
  \enquote{\bibinfo {title} {{A witness for coherent electronic vs
  vibronic-only oscillations in ultrafast spectroscopy}},}\ }%
  \bibfield{journal}{%
  \bibinfo {journal} {J. Chem. Phys.}\ }%
  \textbf{\bibinfo {volume} {136}},\ \bibinfo {pages} {234501} (\bibinfo {year}
  {2012})\BibitemShut{NoStop}%
\bibitem{CaycedoSoler:2012ib}%
  \BibitemOpen
  \bibfield{author}{%
  \bibinfo {author} {\bibfnamefont{Felipe}\ \bibnamefont{Caycedo-Soler}},
  \bibinfo {author} {\bibfnamefont{Alex~W}\ \bibnamefont{Chin}}, \bibinfo
  {author} {\bibfnamefont{Javier}\ \bibnamefont{Almeida}}, \bibinfo {author}
  {\bibfnamefont{Susana~F}\ \bibnamefont{Huelga}},\ and\ \bibinfo {author}
  {\bibfnamefont{Martin~B}\ \bibnamefont{Plenio}},\ }%
  \bibfield{title}{%
  \enquote{\bibinfo {title} {{The nature of the low energy band of the
  Fenna-Matthews-Olson complex: Vibronic signatures}},}\ }%
  \bibfield{journal}{%
  \bibinfo {journal} {J. Chem. Phys.}\ }%
  \textbf{\bibinfo {volume} {136}},\ \bibinfo {pages} {155102} (\bibinfo {year}
  {2012})\BibitemShut{NoStop}%
\bibitem{Butkus:2012hn}%
  \BibitemOpen
  \bibfield{author}{%
  \bibinfo {author} {\bibfnamefont{Vytautas}\ \bibnamefont{Butkus}}, \bibinfo
  {author} {\bibfnamefont{Donatas}\ \bibnamefont{Zigmantas}}, \bibinfo {author}
  {\bibfnamefont{Leonas}\ \bibnamefont{Valkunas}},\ and\ \bibinfo {author}
  {\bibfnamefont{Darius}\ \bibnamefont{Abramavicius}},\ }%
  \bibfield{title}{%
  \enquote{\bibinfo {title} {{Vibrational vs. electronic coherences in 2D
  spectrum of molecular systems}},}\ }%
  \bibfield{journal}{%
  \bibinfo {journal} {Chem. Phys. Lett.}\ }%
  \textbf{\bibinfo {volume} {545}},\ \bibinfo {pages} {40--43} (\bibinfo {year}
  {2012})\BibitemShut{NoStop}%
\bibitem{Butkus:2013fy}%
  \BibitemOpen
  \bibfield{author}{%
  \bibinfo {author} {\bibfnamefont{Vytautas}\ \bibnamefont{Butkus}}, \bibinfo
  {author} {\bibfnamefont{Donatas}\ \bibnamefont{Zigmantas}}, \bibinfo {author}
  {\bibfnamefont{Darius}\ \bibnamefont{Abramavicius}},\ and\ \bibinfo {author}
  {\bibfnamefont{Leonas}\ \bibnamefont{Valkunas}},\ }%
  \bibfield{title}{%
  \enquote{\bibinfo {title} {{Distinctive character of electronic and
  vibrational coherences in disordered molecular aggregates}},}\ }%
  \bibfield{journal}{%
  \bibinfo {journal} {Chem. Phys. Lett.}\ }%
  \textbf{\bibinfo {volume} {587}},\ \bibinfo {pages} {93--98} (\bibinfo {year}
  {2013})\BibitemShut{NoStop}%
\bibitem{Tiwari:2013dt}%
  \BibitemOpen
  \bibfield{author}{%
  \bibinfo {author} {\bibfnamefont{Vivek}\ \bibnamefont{Tiwari}}, \bibinfo
  {author} {\bibfnamefont{William~K}\ \bibnamefont{Peters}},\ and\ \bibinfo
  {author} {\bibfnamefont{David~M}\ \bibnamefont{Jonas}},\ }%
  \bibfield{title}{%
  \enquote{\bibinfo {title} {{Electronic resonance with anticorrelated pigment
  vibrations drives photosynthetic energy transfer outside the adiabatic
  framework}},}\ }%
  \bibfield{journal}{%
  \bibinfo {journal} {Proc. Natl. Acad. Sci. USA}\ }%
  \textbf{\bibinfo {volume} {110}},\ \bibinfo {pages} {1203--1208} (\bibinfo
  {year} {2013})\BibitemShut{NoStop}%
\bibitem{Kreisbeck:2013jva}%
  \BibitemOpen
  \bibfield{author}{%
  \bibinfo {author} {\bibfnamefont{Christoph}\ \bibnamefont{Kreisbeck}},
  \bibinfo {author} {\bibfnamefont{Tobias}\ \bibnamefont{Kramer}},\ and\
  \bibinfo {author} {\bibfnamefont{Al{\'a}n}\ \bibnamefont{Aspuru-Guzik}},\ }%
  \bibfield{title}{%
  \enquote{\bibinfo {title} {{Disentangling Electronic and Vibronic Coherences
  in Two-Dimensional Echo Spectra}},}\ }%
  \bibfield{journal}{%
  \bibinfo {journal} {J. Phys. Chem. B}\ }%
  \textbf{\bibinfo {volume} {117}},\ \bibinfo {pages} {9380--9385} (\bibinfo
  {year} {2013})\BibitemShut{NoStop}%
\bibitem{Chin:2013ia}%
  \BibitemOpen
  \bibfield{author}{%
  \bibinfo {author} {\bibfnamefont{A~W}\ \bibnamefont{Chin}}, \bibinfo {author}
  {\bibfnamefont{J}~\bibnamefont{Prior}}, \bibinfo {author}
  {\bibfnamefont{R}~\bibnamefont{Rosenbach}}, \bibinfo {author}
  {\bibfnamefont{F}~\bibnamefont{Caycedo-Soler}}, \bibinfo {author}
  {\bibfnamefont{S~F}\ \bibnamefont{Huelga}},\ and\ \bibinfo {author}
  {\bibfnamefont{M~B}\ \bibnamefont{Plenio}},\ }%
  \bibfield{title}{%
  \enquote{\bibinfo {title} {{The role of non-equilibrium vibrational
  structures in electronic coherence and recoherence in pigment-protein
  complexes}},}\ }%
  \bibfield{journal}{%
  \bibinfo {journal} {Nat. Phys.}\ }%
  \textbf{\bibinfo {volume} {9}},\ \bibinfo {pages} {113--118} (\bibinfo {year}
  {2013})\BibitemShut{NoStop}%
\bibitem{Chenu:2013cf}%
  \BibitemOpen
  \bibfield{author}{%
  \bibinfo {author} {\bibfnamefont{Aur{\'e}lia}\ \bibnamefont{Chenu}}, \bibinfo
  {author} {\bibfnamefont{Niklas}\ \bibnamefont{Christensson}}, \bibinfo
  {author} {\bibfnamefont{Harald~F}\ \bibnamefont{Kauffmann}},\ and\ \bibinfo
  {author} {\bibfnamefont{Tom{\'a}{\v s}}\ \bibnamefont{Man{\v c}al}},\ }%
  \bibfield{title}{%
  \enquote{\bibinfo {title} {{Enhancement of Vibronic and Ground-State
  Vibrational Coherences in 2D Spectra of Photosynthetic Complexes}},}\ }%
  \bibfield{journal}{%
  \bibinfo {journal} {Sci. Rep.}\ }%
  \textbf{\bibinfo {volume} {3}},\ \bibinfo {pages} {2029} (\bibinfo {year}
  {2013})\BibitemShut{NoStop}%
\bibitem{Plenio:2013bg}%
  \BibitemOpen
  \bibfield{author}{%
  \bibinfo {author} {\bibfnamefont{M~B}\ \bibnamefont{Plenio}}, \bibinfo
  {author} {\bibfnamefont{J}~\bibnamefont{Almeida}},\ and\ \bibinfo {author}
  {\bibfnamefont{S~F}\ \bibnamefont{Huelga}},\ }%
  \bibfield{title}{%
  \enquote{\bibinfo {title} {{Origin of long-lived oscillations in 2D-spectra
  of a quantum vibronic model: Electronic versus vibrational coherence}},}\ }%
  \bibfield{journal}{%
  \bibinfo {journal} {J. Chem. Phys.}\ }%
  \textbf{\bibinfo {volume} {139}},\ \bibinfo {pages} {235102} (\bibinfo {year}
  {2013})\BibitemShut{NoStop}%
\bibitem{Rivera:2013kb}%
  \BibitemOpen
  \bibfield{author}{%
  \bibinfo {author} {\bibfnamefont{Eva}\ \bibnamefont{Rivera}}, \bibinfo
  {author} {\bibfnamefont{Daniel}\ \bibnamefont{Montemayor}}, \bibinfo {author}
  {\bibfnamefont{Marco}\ \bibnamefont{Masia}},\ and\ \bibinfo {author}
  {\bibfnamefont{David~F}\ \bibnamefont{Coker}},\ }%
  \bibfield{title}{%
  \enquote{\bibinfo {title} {{Influence of Site-Dependent
  Pigment{\textendash}Protein Interactions on Excitation Energy Transfer in
  Photosynthetic Light Harvesting}},}\ }%
  \bibfield{journal}{%
  \bibinfo {journal} {J. Phys. Chem. B}\ }%
  \textbf{\bibinfo {volume} {117}},\ \bibinfo {pages} {5510--5521} (\bibinfo
  {year} {2013})\BibitemShut{NoStop}%
\bibitem{Seibt:2013dp}%
  \BibitemOpen
  \bibfield{author}{%
  \bibinfo {author} {\bibfnamefont{Joachim}\ \bibnamefont{Seibt}}\ and\
  \bibinfo {author} {\bibfnamefont{T{\~o}nu}\ \bibnamefont{Pullerits}},\ }%
  \bibfield{title}{%
  \enquote{\bibinfo {title} {{Beating Signals in 2D Spectroscopy: Electronic or
  Nuclear Coherences? Application to a Quantum Dot Model System}},}\ }%
  \bibfield{journal}{%
  \bibinfo {journal} {J. Phys. Chem. C}\ }%
  \textbf{\bibinfo {volume} {117}},\ \bibinfo {pages} {18728--18737} (\bibinfo
  {year} {2013})\BibitemShut{NoStop}%
\bibitem{Halpin:2014jd}%
  \BibitemOpen
  \bibfield{author}{%
  \bibinfo {author} {\bibfnamefont{Alexei}\ \bibnamefont{Halpin}}, \bibinfo
  {author} {\bibfnamefont{Philip J~M}\ \bibnamefont{Johnson}}, \bibinfo
  {author} {\bibfnamefont{Roel}\ \bibnamefont{Tempelaar}}, \bibinfo {author}
  {\bibfnamefont{R~Scott}\ \bibnamefont{Murphy}}, \bibinfo {author}
  {\bibfnamefont{Jasper}\ \bibnamefont{Knoester}}, \bibinfo {author}
  {\bibfnamefont{Thomas L~C}\ \bibnamefont{Jansen}},\ and\ \bibinfo {author}
  {\bibfnamefont{R~J~Dwayne}\ \bibnamefont{Miller}},\ }%
  \bibfield{title}{%
  \enquote{\bibinfo {title} {{Two-dimensional spectroscopy of a molecular dimer
  unveils the effects of vibronic coupling on exciton coherences}},}\ }%
  \bibfield{journal}{%
  \bibinfo {journal} {Nat. Chem.}\ }%
  \textbf{\bibinfo {volume} {6}},\ \bibinfo {pages} {196--201} (\bibinfo {year}
  {2014})\BibitemShut{NoStop}%
\bibitem{Tempelaar:2014vu}%
  \BibitemOpen
  \bibfield{author}{%
  \bibinfo {author} {\bibfnamefont{R}~\bibnamefont{Tempelaar}}, \bibinfo
  {author} {\bibfnamefont{Thomas L.~C.}\ \bibnamefont{Jansen}},\ and\ \bibinfo
  {author} {\bibfnamefont{Jasper}\ \bibnamefont{Knoester}},\ }%
  \bibfield{title}{%
  \enquote{\bibinfo {title} {{Vibrational Beatings Conceal Evidence of
  Electronic Coherence in the FMO Light-Harvesting Complex}},}\ }%
  \bibfield{journal}{%
  \bibinfo {journal} {J. Phys. Chem. B}\ }%
  \textbf{\bibinfo {volume} {118}},\ \bibinfo {pages} {12865--12872} (\bibinfo
  {year} {2014})\BibitemShut{NoStop}%
\bibitem{OReilly:2014it}%
  \BibitemOpen
  \bibfield{author}{%
  \bibinfo {author} {\bibfnamefont{Edward~J}\ \bibnamefont{O'Reilly}}\ and\
  \bibinfo {author} {\bibfnamefont{Alexandra}\ \bibnamefont{Olaya-Castro}},\ }%
  \bibfield{title}{%
  \enquote{\bibinfo {title} {{Non-classicality of the molecular vibrations
  assisting exciton energy transfer at room temperature}},}\ }%
  \bibfield{journal}{%
  \bibinfo {journal} {Nat. Commun.}\ }%
  \textbf{\bibinfo {volume} {5}},\ \bibinfo {pages} {1--10} (\bibinfo {year}
  {2014})\BibitemShut{NoStop}%
\bibitem{Perlik:2014bd}%
  \BibitemOpen
  \bibfield{author}{%
  \bibinfo {author} {\bibfnamefont{V{\'a}clav}\ \bibnamefont{Perl{\'\i}k}},
  \bibinfo {author} {\bibfnamefont{Craig}\ \bibnamefont{Lincoln}}, \bibinfo
  {author} {\bibfnamefont{Franti{\v s}ek}\ \bibnamefont{{\v{S}}anda}},\ and\
  \bibinfo {author} {\bibfnamefont{J{\"u}rgen}\ \bibnamefont{Hauer}},\ }%
  \bibfield{title}{%
  \enquote{\bibinfo {title} {{Distinguishing Electronic and Vibronic Coherence
  in 2D Spectra by Their Temperature Dependence}},}\ }%
  \bibfield{journal}{%
  \bibinfo {journal} {J. Phys. Chem. Lett.}\ }%
  \textbf{\bibinfo {volume} {5}},\ \bibinfo {pages} {404--407} (\bibinfo {year}
  {2014})\BibitemShut{NoStop}%
\bibitem{Ishizaki:2009jg}%
  \BibitemOpen
  \bibfield{author}{%
  \bibinfo {author} {\bibfnamefont{Akihito}\ \bibnamefont{Ishizaki}}\ and\
  \bibinfo {author} {\bibfnamefont{Graham~R}\ \bibnamefont{Fleming}},\ }%
  \bibfield{title}{%
  \enquote{\bibinfo {title} {{Unified treatment of quantum coherent and
  incoherent hopping dynamics in electronic energy transfer: reduced hierarchy
  equation approach.}}.}\ }%
  \bibfield{journal}{%
  \bibinfo {journal} {J. Chem. Phys.}\ }%
  \textbf{\bibinfo {volume} {130}},\ \bibinfo {pages} {234111} (\bibinfo {year}
  {2009})\BibitemShut{NoStop}%
\bibitem{Ishizaki:2009ky}%
  \BibitemOpen
  \bibfield{author}{%
  \bibinfo {author} {\bibfnamefont{Akihito}\ \bibnamefont{Ishizaki}}\ and\
  \bibinfo {author} {\bibfnamefont{Graham~R}\ \bibnamefont{Fleming}},\ }%
  \bibfield{title}{%
  \enquote{\bibinfo {title} {{Theoretical examination of quantum coherence in a
  photosynthetic system at physiological temperature.}}.}\ }%
  \bibfield{journal}{%
  \bibinfo {journal} {Proc. Natl. Acad. Sci. USA}\ }%
  \textbf{\bibinfo {volume} {106}},\ \bibinfo {pages} {17255--17260} (\bibinfo
  {year} {2009})\BibitemShut{NoStop}%
\bibitem{Wendling:2000ha}%
  \BibitemOpen
  \bibfield{author}{%
  \bibinfo {author} {\bibfnamefont{Markus}\ \bibnamefont{Wendling}}, \bibinfo
  {author} {\bibfnamefont{T{\~o}nu}\ \bibnamefont{Pullerits}}, \bibinfo
  {author} {\bibfnamefont{Milosz~A}\ \bibnamefont{Przyjalgowski}}, \bibinfo
  {author} {\bibfnamefont{Simone I~E}\ \bibnamefont{Vulto}}, \bibinfo {author}
  {\bibfnamefont{Thijs~J}\ \bibnamefont{Aartsma}}, \bibinfo {author}
  {\bibfnamefont{Rienk}\ \bibnamefont{van Grondelle}},\ and\ \bibinfo {author}
  {\bibfnamefont{Herbert}\ \bibnamefont{van Amerongen}},\ }%
  \bibfield{title}{%
  \enquote{\bibinfo {title} {{Electron-Vibrational Coupling in the
  Fenna-Matthews-Olson Complex of Prosthecochloris a estuarii Determined by
  Temperature-Dependent Absorption and Fluorescence Line-Narrowing
  Measurements}},}\ }%
  \bibfield{journal}{%
  \bibinfo {journal} {J. Phys. Chem. B}\ }%
  \textbf{\bibinfo {volume} {104}},\ \bibinfo {pages} {5825--5831} (\bibinfo
  {year} {2000})\BibitemShut{NoStop}%
\bibitem{Ratsep:2007fq}%
  \BibitemOpen
  \bibfield{author}{%
  \bibinfo {author} {\bibfnamefont{Margus}\ \bibnamefont{R{\"a}tsep}}\ and\
  \bibinfo {author} {\bibfnamefont{Arvi}\ \bibnamefont{Freiberg}},\ }%
  \bibfield{title}{%
  \enquote{\bibinfo {title} {{Electron-phonon and vibronic couplings in the FMO
  bacteriochlorophyll {\it a} antenna complex studied by difference
  fluorescence line narrowing}},}\ }%
  \bibfield{journal}{%
  \bibinfo {journal} {J. Lumin.}\ }%
  \textbf{\bibinfo {volume} {127}},\ \bibinfo {pages} {251--259} (\bibinfo
  {year} {2007})\BibitemShut{NoStop}%
\bibitem{Ratsep:2011cq}%
  \BibitemOpen
  \bibfield{author}{%
  \bibinfo {author} {\bibfnamefont{Margus}\ \bibnamefont{R{\"a}tsep}}, \bibinfo
  {author} {\bibfnamefont{Zheng-Li}\ \bibnamefont{Cai}}, \bibinfo {author}
  {\bibfnamefont{Jeffrey~R}\ \bibnamefont{Reimers}},\ and\ \bibinfo {author}
  {\bibfnamefont{Arvi}\ \bibnamefont{Freiberg}},\ }%
  \bibfield{title}{%
  \enquote{\bibinfo {title} {{Demonstration and interpretation of significant
  asymmetry in the low-resolution and high-resolution $Q_y$ fluorescence and
  absorption spectra of bacteriochlorophyll {\it a}}},}\ }%
  \bibfield{journal}{%
  \bibinfo {journal} {J. Chem. Phys.}\ }%
  \textbf{\bibinfo {volume} {134}},\ \bibinfo {pages} {024506} (\bibinfo {year}
  {2011})\BibitemShut{NoStop}%
\bibitem{Schulze:2014iv}%
  \BibitemOpen
  \bibfield{author}{%
  \bibinfo {author} {\bibfnamefont{Jan}\ \bibnamefont{Schulze}}, \bibinfo
  {author} {\bibfnamefont{Magne}\ \bibnamefont{Torbj{\"o}rnsson}}, \bibinfo
  {author} {\bibfnamefont{Oliver}\ \bibnamefont{K{\"u}hn}},\ and\ \bibinfo
  {author} {\bibfnamefont{T{\~o}nu}\ \bibnamefont{Pullerits}},\ }%
  \bibfield{title}{%
  \enquote{\bibinfo {title} {{Exciton coupling induces vibronic hyperchromism
  in light-harvesting complexes}},}\ }%
  \bibfield{journal}{%
  \bibinfo {journal} {New J. Phys.}\ }%
  \textbf{\bibinfo {volume} {16}},\ \bibinfo {pages} {045010} (\bibinfo {year}
  {2014})\BibitemShut{NoStop}%
\bibitem{Ishizaki:2010ft}%
  \BibitemOpen
  \bibfield{author}{%
  \bibinfo {author} {\bibfnamefont{Akihito}\ \bibnamefont{Ishizaki}}\ and\
  \bibinfo {author} {\bibfnamefont{Graham~R}\ \bibnamefont{Fleming}},\ }%
  \bibfield{title}{%
  \enquote{\bibinfo {title} {{Quantum superpositions in photosynthetic light
  harvesting: delocalization and entanglement}},}\ }%
  \bibfield{journal}{%
  \bibinfo {journal} {New J. Phys.}\ }%
  \textbf{\bibinfo {volume} {12}},\ \bibinfo {pages} {055004} (\bibinfo {year}
  {2010})\BibitemShut{NoStop}%
\bibitem{Ishizaki:2013jg}%
  \BibitemOpen
  \bibfield{author}{%
  \bibinfo {author} {\bibfnamefont{Akihito}\ \bibnamefont{Ishizaki}},\ }%
  \bibfield{title}{%
  \enquote{\bibinfo {title} {{Interactions between Quantum Mixing and the
  Environmental Dynamics Controlling Ultrafast Photoinduced Electron Transfer
  and Its Temperature Dependence}},}\ }%
  \bibfield{journal}{%
  \bibinfo {journal} {Chem. Lett.}\ }%
  \textbf{\bibinfo {volume} {42}},\ \bibinfo {pages} {1406--1408} (\bibinfo
  {year} {2013})\BibitemShut{NoStop}%
\bibitem{Fujihashi:2015kz}%
  \BibitemOpen
  \bibfield{author}{%
  \bibinfo {author} {\bibfnamefont{Yuta}\ \bibnamefont{Fujihashi}}, \bibinfo
  {author} {\bibfnamefont{Graham~R}\ \bibnamefont{Fleming}},\ and\ \bibinfo
  {author} {\bibfnamefont{Akihito}\ \bibnamefont{Ishizaki}},\ }%
  \bibfield{title}{%
  \enquote{\bibinfo {title} {{Impact of environmentally induced fluctuations on
  quantum mechanically mixed electronic and vibrational pigment states in
  photosynthetic energy transfer and 2D electronic spectra}},}\ }%
  \bibfield{journal}{%
  \bibinfo {journal} {J. Chem. Phys.}\ }%
  \textbf{\bibinfo {volume} {142}},\ \bibinfo {pages} {212403} (\bibinfo {year}
  {2015}),\
  \Eprint{http://arxiv.org/abs/1505.05281}{arXiv:1505.05281}\BibitemShut{NoSto%
p}%
%%CITATION = ARXIV:1505.05281;%%
\bibitem{vanAmerongen:2000fy}%
  \BibitemOpen
  \bibfield{author}{%
  \bibinfo {author} {\bibfnamefont{H}~\bibnamefont{van Amerongen}}, \bibinfo
  {author} {\bibfnamefont{L}~\bibnamefont{Valkunas}},\ and\ \bibinfo {author}
  {\bibfnamefont{R}~\bibnamefont{van Grondelle}},\ }%
  \emph{\bibinfo {title} {{Photosynthetic excitons}}}\ (\bibinfo {publisher}
  {World Scientific},\ \bibinfo {address} {Singapore},\ \bibinfo {year}
  {2000})\BibitemShut{NoStop}%
\bibitem{Cho:2005bm}%
  \BibitemOpen
  \bibfield{author}{%
  \bibinfo {author} {\bibfnamefont{Minhaeng}\ \bibnamefont{Cho}}, \bibinfo
  {author} {\bibfnamefont{Harsha~M}\ \bibnamefont{Vaswani}}, \bibinfo {author}
  {\bibfnamefont{Tobias}\ \bibnamefont{Brixner}}, \bibinfo {author}
  {\bibfnamefont{Jens}\ \bibnamefont{Stenger}},\ and\ \bibinfo {author}
  {\bibfnamefont{Graham~R}\ \bibnamefont{Fleming}},\ }%
  \bibfield{title}{%
  \enquote{\bibinfo {title} {{Exciton Analysis in 2D Electronic
  Spectroscopy}},}\ }%
  \bibfield{journal}{%
  \bibinfo {journal} {J. Phys. Chem. B}\ }%
  \textbf{\bibinfo {volume} {109}},\ \bibinfo {pages} {10542--10556} (\bibinfo
  {year} {2005})\BibitemShut{NoStop}%
\bibitem{Adolphs:2006ey}%
  \BibitemOpen
  \bibfield{author}{%
  \bibinfo {author} {\bibfnamefont{Julia}\ \bibnamefont{Adolphs}}\ and\
  \bibinfo {author} {\bibfnamefont{Thomas}\ \bibnamefont{Renger}},\ }%
  \bibfield{title}{%
  \enquote{\bibinfo {title} {{How Proteins Trigger Excitation Energy Transfer
  in the FMO Complex of Green Sulfur Bacteria}},}\ }%
  \bibfield{journal}{%
  \bibinfo {journal} {Biophys. J.}\ }%
  \textbf{\bibinfo {volume} {91}},\ \bibinfo {pages} {2778--2797} (\bibinfo
  {year} {2006})\BibitemShut{NoStop}%
\bibitem{Brixner:2004ko}%
  \BibitemOpen
  \bibfield{author}{%
  \bibinfo {author} {\bibfnamefont{Tobias}\ \bibnamefont{Brixner}}, \bibinfo
  {author} {\bibfnamefont{Tom{\'a}{\v s}}\ \bibnamefont{Man{\v c}al}}, \bibinfo
  {author} {\bibfnamefont{Igor~V}\ \bibnamefont{Stiopkin}},\ and\ \bibinfo
  {author} {\bibfnamefont{Graham~R}\ \bibnamefont{Fleming}},\ }%
  \bibfield{title}{%
  \enquote{\bibinfo {title} {{Phase-stabilized two-dimensional electronic
  spectroscopy}},}\ }%
  \bibfield{journal}{%
  \bibinfo {journal} {J. Chem. Phys.}\ }%
  \textbf{\bibinfo {volume} {121}},\ \bibinfo {pages} {4221} (\bibinfo {year}
  {2004})\BibitemShut{NoStop}%
\bibitem{Bruggemann:2007gu}%
  \BibitemOpen
  \bibfield{author}{%
  \bibinfo {author} {\bibfnamefont{Ben}\ \bibnamefont{Br{\"u}ggemann}},
  \bibinfo {author} {\bibfnamefont{P{\"a}r}\ \bibnamefont{Kjellberg}},\ and\
  \bibinfo {author} {\bibfnamefont{T{\~o}nu}\ \bibnamefont{Pullerits}},\ }%
  \bibfield{title}{%
  \enquote{\bibinfo {title} {{Non-perturbative calculation of 2D spectra in
  heterogeneous systems: Exciton relaxation in the FMO complex}},}\ }%
  \bibfield{journal}{%
  \bibinfo {journal} {Chem. Phys. Lett.}\ }%
  \textbf{\bibinfo {volume} {444}},\ \bibinfo {pages} {192--196} (\bibinfo
  {year} {2007})\BibitemShut{NoStop}%
\bibitem{Kubo:1985bs}%
  \BibitemOpen
  \bibfield{author}{%
  \bibinfo {author} {\bibfnamefont{Ryogo}\ \bibnamefont{Kubo}}, \bibinfo
  {author} {\bibfnamefont{Morikazu}\ \bibnamefont{Toda}},\ and\ \bibinfo
  {author} {\bibfnamefont{Natsuki}\ \bibnamefont{Hashitsume}},\ }%
  \emph{\bibinfo {title} {{Statistical Physics II}}}\ (\bibinfo {publisher}
  {Springer},\ \bibinfo {address} {Berlin, Heidelberg},\ \bibinfo {year}
  {1985})\BibitemShut{NoStop}%
\bibitem{Womick:2011iw}%
  \BibitemOpen
  \bibfield{author}{%
  \bibinfo {author} {\bibfnamefont{Jordan~M}\ \bibnamefont{Womick}}\ and\
  \bibinfo {author} {\bibfnamefont{Andrew~M}\ \bibnamefont{Moran}},\ }%
  \bibfield{title}{%
  \enquote{\bibinfo {title} {{Vibronic Enhancement of Exciton Sizes and Energy
  Transport in Photosynthetic Complexes}},}\ }%
  \bibfield{journal}{%
  \bibinfo {journal} {J. Phys. Chem. B}\ }%
  \textbf{\bibinfo {volume} {115}},\ \bibinfo {pages} {1347--1356} (\bibinfo
  {year} {2011})\BibitemShut{NoStop}%
\bibitem{Grabert:1988ha}%
  \BibitemOpen
  \bibfield{author}{%
  \bibinfo {author} {\bibfnamefont{H}~\bibnamefont{Grabert}}, \bibinfo {author}
  {\bibfnamefont{P}~\bibnamefont{Schramm}},\ and\ \bibinfo {author}
  {\bibfnamefont{G~L}\ \bibnamefont{Ingold}},\ }%
  \bibfield{title}{%
  \enquote{\bibinfo {title} {{Quantum Brownian motion: the functional integral
  approach}},}\ }%
  \bibfield{journal}{%
  \bibinfo {journal} {Phys. Rep.}\ }%
  \textbf{\bibinfo {volume} {168}},\ \bibinfo {pages} {115--207} (\bibinfo
  {year} {1988})\BibitemShut{NoStop}%
\bibitem{Tanimura:2014er}%
  \BibitemOpen
  \bibfield{author}{%
  \bibinfo {author} {\bibfnamefont{Yoshitaka}\ \bibnamefont{Tanimura}},\ }%
  \bibfield{title}{%
  \enquote{\bibinfo {title} {{Reduced hierarchical equations of motion in real
  and imaginary time: Correlated initial states and thermodynamic
  quantities}},}\ }%
  \bibfield{journal}{%
  \bibinfo {journal} {J. Chem. Phys.}\ }%
  \textbf{\bibinfo {volume} {141}},\ \bibinfo {pages} {044114} (\bibinfo {year}
  {2014})\BibitemShut{NoStop}%
\bibitem{Ishizaki:2010fx}%
  \BibitemOpen
  \bibfield{author}{%
  \bibinfo {author} {\bibfnamefont{Akihito}\ \bibnamefont{Ishizaki}}, \bibinfo
  {author} {\bibfnamefont{Tessa~R}\ \bibnamefont{Calhoun}}, \bibinfo {author}
  {\bibfnamefont{Gabriela~S}\ \bibnamefont{Schlau-Cohen}},\ and\ \bibinfo
  {author} {\bibfnamefont{Graham~R}\ \bibnamefont{Fleming}},\ }%
  \bibfield{title}{%
  \enquote{\bibinfo {title} {{Quantum coherence and its interplay with protein
  environments in photosynthetic electronic energy transfer}},}\ }%
  \bibfield{journal}{%
  \bibinfo {journal} {Phys. Chem. Chem. Phys.}\ }%
  \textbf{\bibinfo {volume} {12}},\ \bibinfo {pages} {7319} (\bibinfo {year}
  {2010})\BibitemShut{NoStop}%
\bibitem{Tronrud:2009ch}%
  \BibitemOpen
  \bibfield{author}{%
  \bibinfo {author} {\bibfnamefont{Dale~E}\ \bibnamefont{Tronrud}}, \bibinfo
  {author} {\bibfnamefont{Jianzhong}\ \bibnamefont{Wen}}, \bibinfo {author}
  {\bibfnamefont{Leslie}\ \bibnamefont{Gay}},\ and\ \bibinfo {author}
  {\bibfnamefont{Robert~E}\ \bibnamefont{Blankenship}},\ }%
  \bibfield{title}{%
  \enquote{\bibinfo {title} {{The structural basis for the difference in
  absorbance spectra for the FMO antenna protein from various green sulfur
  bacteria}},}\ }%
  \bibfield{journal}{%
  \bibinfo {journal} {Photosynth. Res.}\ }%
  \textbf{\bibinfo {volume} {100}},\ \bibinfo {pages} {79--87} (\bibinfo {year}
  {2009})\BibitemShut{NoStop}%
\bibitem{Fransted:2012fs}%
  \BibitemOpen
  \bibfield{author}{%
  \bibinfo {author} {\bibfnamefont{Kelly~A}\ \bibnamefont{Fransted}}, \bibinfo
  {author} {\bibfnamefont{Justin~R}\ \bibnamefont{Caram}}, \bibinfo {author}
  {\bibfnamefont{Dugan}\ \bibnamefont{Hayes}},\ and\ \bibinfo {author}
  {\bibfnamefont{Gregory~S}\ \bibnamefont{Engel}},\ }%
  \bibfield{title}{%
  \enquote{\bibinfo {title} {{Two-dimensional electronic spectroscopy of
  bacteriochlorophyll a in solution: Elucidating the coherence dynamics of the
  Fenna-Matthews-Olson complex using its chromophore as a control}},}\ }%
  \bibfield{journal}{%
  \bibinfo {journal} {J. Chem. Phys.}\ }%
  \textbf{\bibinfo {volume} {137}},\ \bibinfo {pages} {125101} (\bibinfo {year}
  {2012})\BibitemShut{NoStop}%
\end{thebibliography}%
%%%%%%%%%%%%%%%%%%%%%%%%%%%%%%%%%%%%%%%%%%%%%%%%%%%%%%%%%%%%%%%%%%%%%
\newpage

%%%%%%%%%%%%

\newpage

%%%%%%%%%%%%

\newpage

%%%%%%%%%%%%

\newpage

%%%%%%%%%%%%

\end{document}